\documentclass[twocolumn,floatfix,aps,superscriptaddress,prb]{revtex4-2}
\usepackage{graphicx}
\usepackage{amsmath}
\usepackage{amssymb}
\usepackage{bm}
\usepackage{hyperref}

\usepackage{color}

\begin{document}

\title{Breathing mode in open-orbit magnetotransport:\\ 
a magnetic lens with a quantum mechanical focal length}
\author{D. O. Oriekhov}
\affiliation{Instituut-Lorentz, Universiteit Leiden, P.O. Box 9506, 2300 RA Leiden, The Netherlands}
\author{T. T. Osterholt}
\affiliation{Instituut-Lorentz, Universiteit Leiden, P.O. Box 9506, 2300 RA Leiden, The Netherlands}
\author{T. Vakhtel}
\affiliation{Instituut-Lorentz, Universiteit Leiden, P.O. Box 9506, 2300 RA Leiden, The Netherlands}
\author{A. R. Akhmerov}
\affiliation{Kavli Institute of Nanoscience, Delft University of Technology, P.O. Box 4056, 2600 GA Delft, The Netherlands}
\author{C. W. J. Beenakker}
\affiliation{Instituut-Lorentz, Universiteit Leiden, P.O. Box 9506, 2300 RA Leiden, The Netherlands}
\date{July 2022}
\begin{abstract}
We consider the propagation of electrons in a lattice with an anisotropic dispersion in the $x$--$y$ plane (lattice constant $a$), such that it supports open orbits along the $x$-axis in an out-of-plane magnetic field $B$.  We show that a point source excites a ``breathing mode'', a state that periodically spreads out and refocuses after having propagated over a distance $\ell =(eaB/h)^{-1}$ in the $x$-direction. Unlike known magnetic focusing effects, governed by the classical cyclotron radius, this is an intrinsically quantum mechanical effect with a focal length $\propto\hbar$.
\end{abstract}
\maketitle

\section{Introduction}
\label{intro}

The Lorentz force from a magnetic field may act as a lens for electrons, by focusing their trajectories down to a point of size limited only by their wave length. In the solid state such electron optics was pioneered half a century ago by Sharvin and Tsoi  \cite{Sha65a,Sha65b,Tso74}, enabled by the availability of single crystals with mean free paths of several millimeters --- well above the typical focal lengths of the magnetic lens. Geometric optics is sufficient in metals \cite{Tso99,Boz14}, in semiconductors and in graphene the larger wave length introduces diffraction and interference effects \cite{Hou89,Rak10,Ste13,Tay13}.

Irrespective of these quantum effects, the magnetic focusing itself is still an essentially classical effect --- the focal length is given by the classical cyclotron radius $p_{\rm F}/eB$ (ratio of Fermi momentum and magnetic field). In what follows we will describe a magnetic focusing effect that is intrinsically quantum mechanical. The focusing mechanism is Bragg reflection at Brillouin zone boundaries, resulting in a parametrically larger focal length, with Fermi momentum $p_{\rm F}$ replaced by the Bragg momentum transfer $\hbar/a$ (inverse lattice constant).  

We build on our recent study of magnetotransport in twisted bilayer graphene \cite{Vak22}, where a precise mathematical mapping was found onto Bloch oscillations in an electric quantum walk \cite{Ced13,Arn20,Ced21}. The mapping of space onto time and magnetic field onto electric field was shown to result in a ``breathing mode'' \cite{Kur96,Har04}, a wave function that periodically expands and contracts. The mapping relied on the special nature of the scattering problem in the graphene bilayer \cite{Beu20}, where electrons propagate in topologically protected chiral modes on a triangular network of domain walls \cite{San13,Efi18}. 

Here we take a broader perspective and develop a general theory for breathing modes that applies to any band structure which supports open orbits in a magnetic field. It applies in particular to layered materials with a strongly anisotropic dispersion, of recent interest in this context \cite{Put20,Vil20}. We present both a fully quantum mechanical calculation and a semiclassical description of the breathing mode, and test this by comparing with computer simulations of a tight-binding model.

\section{Calculation of the breathing mode}
\label{sec_breathing}

\begin{figure}[tb]
\centerline{\includegraphics[width=0.9\linewidth]{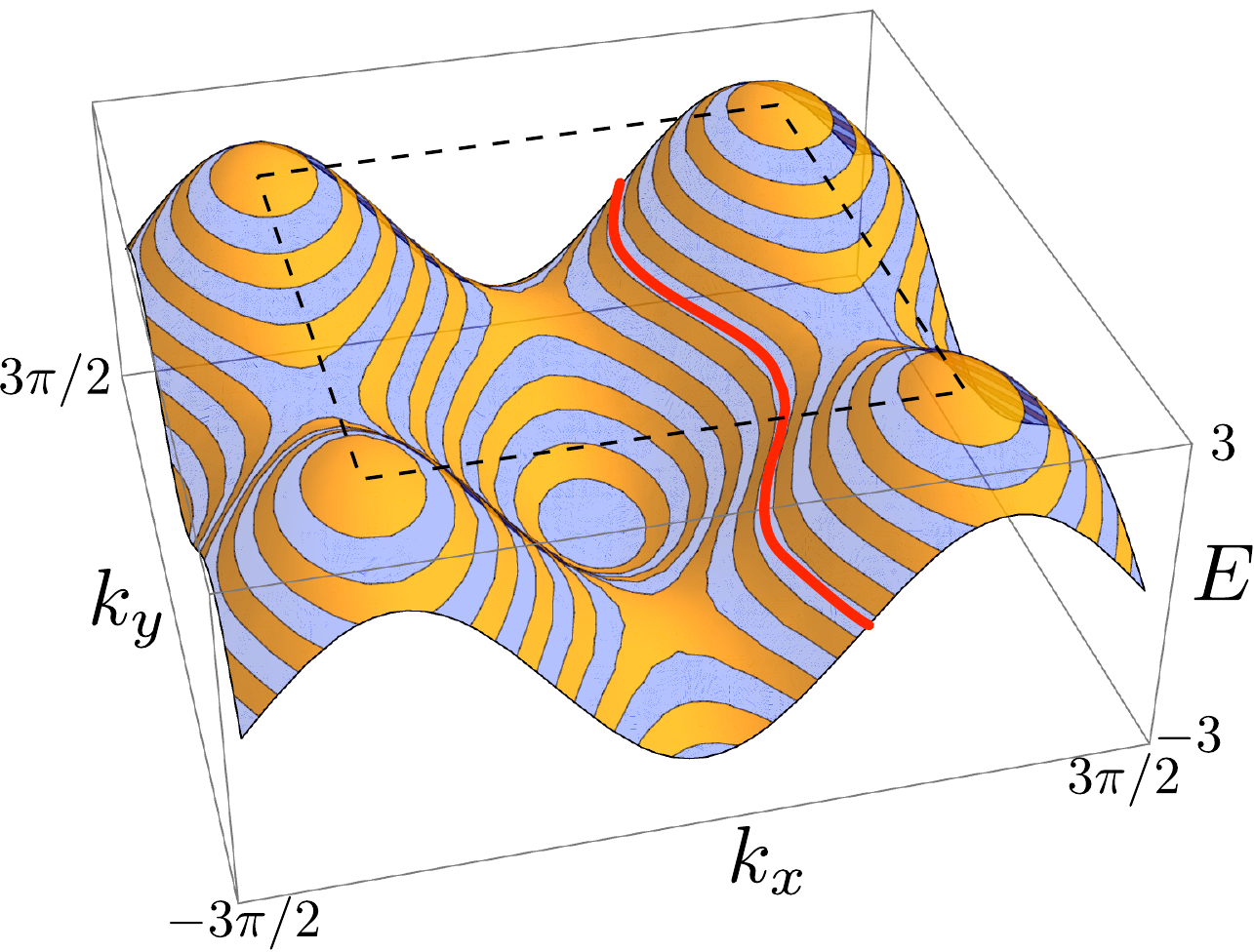}}
\caption{Equi-energy contours of the 2D dispersion $E(k_x,k_y) = -2\cos k_x-\cos k_y$ (dimensionless units). The black dotted square indicates the Brillouin zone, the red curve is an open orbit at the Fermi energy $E_{\rm F}=0$, given by $k_x+\varepsilon(k_y)=0$, with $\varepsilon(k_y)=-\arccos(-\tfrac{1}{2}\cos k_y)$.
}
\label{fig_dispersion}
\end{figure}

An open orbit in the Brillouin zone is an equi-energy contour that crosses the Brillouin zone boundaries. In the repeated zone scheme it therefore runs through the whole reciprocal space, without closing on itself. The open orbits in a plane perpendicular to an applied magnetic field govern the electrical transport properties. We orient the field in the $z$-direction and focus on an open orbit in the $x$--$y$ plane. An example on the 2D square lattice is shown in Fig.\ \ref{fig_dispersion}.

As an effective low-energy description of an open orbit we consider a two-dimensional (2D) Bloch band near the Fermi energy $E_{\rm F}=0$ in the first Brillouin zone, described by the Hamiltonian
\begin{equation}
H=\hbar v_x k_x+\varepsilon(k_y).
\end{equation}
The momentum operator is $\bm{k}=-i\partial/\partial\bm{r}$. The open orbit has the equi-energy contour $\varepsilon(k_y)+\hbar v_x k_x=0$, with $\varepsilon(k_y)=\varepsilon(k_y+2\pi/a_y)$ for lattice constant $a_y$.

The vector potential is introduced via the substitution $\hbar\bm{k}\mapsto \hbar\bm{k}-e\bm{A}$ (taking the electron charge as $+e$). We choose the gauge $\bm{A}=(-yB(x),0,0)$, corresponding to the magnetic field $\bm{B}=\bigl(0,0,B(x)\bigr)$. We will later specialize to the case $B(x)=B_0$ of a constant field.

Eigenstates $\Psi(x,k_y)$ of $H$ at energy $E=0$, in a mixed coordinate-momentum representation, satisfy
\begin{equation}
iv_x(-\hbar\partial_x+eB(x)\partial_{k_y})\Psi(x,k_y)=-\varepsilon(k_y)\Psi(x,k_y).\label{contevolution}
\end{equation}
A similar partial differential equation has been studied in the context of Wannier-Stark localization \cite{Zha95}, and we can adapt that method of solution.
 
We define the field integral
\begin{equation}
C(x)=\int_0^x B(x')\,dx',
\end{equation}
and note that 
\begin{equation}
\hbar\partial_x f\bigl( \hbar k_y+eC(x)\bigr)=eB(x)\partial_{k_y}f\bigl( \hbar k_y+eC(x)\bigr), 
\end{equation}
for any function $f$. We thus find the solution
\begin{align}
&\Psi(x,k_y)=\Psi\bigl(0, k_y+(e/\hbar)C(x)\bigr)\exp\bigl(-i\omega(x,k_y)\bigr),\label{psixky}\\
&\omega(x,k_y)=\int_0^x \frac{dx'}{\hbar v_x} \,\varepsilon\bigl(k_y+(e/\hbar)C(x)-(e/\hbar)C(x')\bigr).
\end{align}
For an initial condition $\Psi(0,k_y)\equiv 1$ that is localized at $y=0$ we obtain the real space profile
\begin{equation}
\psi(x,y)=a_y\int_0^{2\pi/a_y}\frac{dk_y}{2\pi} e^{iyk_y}\exp\bigl(-i\omega(x,k_y)\bigr).\label{Psixmresult}
\end{equation}

The first moment of the transverse displacement vanishes,
\begin{align}
\langle y\rangle_x&=a_y\sum_{m=-\infty}^\infty m|\psi(x,m a_y)|^2\nonumber\\
&=i a_y\int_0^{2\pi/a_y}\frac{dk_y}{2\pi}\,\Psi^\ast(x,k_y) \partial_{k_y}\Psi(x,k_y)\nonumber\\
&=a_y\int_0^{2\pi/a_y}\frac{dk_y}{2\pi}\,\partial_{k_y}\omega(x,k_y)=0.
\end{align}
The second moment is given by
\begin{align}
\langle y^2\rangle_x={}&a_y^2\sum_{m=-\infty}^\infty m^2|\psi(x,m a_y)|^2\nonumber\\
={}&a_y\int_0^{2\pi/a_y}\frac{dk_y}{2\pi}|\partial_{k_y}\Psi(x,k_y)|^2\nonumber\\
={}&a_y\int_0^{2\pi/a_y}\frac{dk_y}{2\pi}\left(\partial_{k_y}\omega(x,k_y)\right)^2.\label{y2exact}
\end{align}

Specializing now to a constant magnetic field, we have $C(x)=B_0x$ and 
\begin{equation}
\omega(x,k_y)=(eB_0 v_x)^{-1}\int_{k_y}^{k_y+eB_0x/\hbar}dq\,\varepsilon(q).
\end{equation}
We conclude that
\begin{equation}
\psi(x+2\pi\hbar/eB_0a_y,y)=\psi(x,y)e^{-i\alpha},
\end{equation}
for some constant phase $\alpha$, so the density $|\psi(x,y)|^2$ is periodic in $x$ with period
\begin{equation}
\ell=\frac{h}{eB_0a_y}=\frac{ a_x\Phi_0}{\Phi}.
\end{equation}
Here $\Phi=Ba_xa_y$ is the flux through a unit cell and $\Phi_0=h/e$ is the flux quantum.

The transverse displacement has variance
\begin{equation}
\langle y^2\rangle_x=\frac{a_y}{(eB_0v_x)^{2}}\int_0^{2\pi/a_y}\frac{dk_y}{2\pi}\bigl(\varepsilon(k_y+eB_0x)-\varepsilon(k_y)\bigr)^2,\label{y2analytics}
\end{equation}
which vanishes when $x=n\ell$, $n=1,2,\ldots$ --- the breathing mode refocuses to a single lattice site.

\section{Tight-binding model}
\label{sec_TB}

We test this analytical theory numerically on the tight-binding model of a 2D square lattice (lattice constant $a_x=a_y=a$) with anisotropic nearest-neighbor hopping energies $t_x$ and $t_y$ in the $x$- and $y$-directions. In the plots we take $t_y/t_x\equiv \tau=1/2$. The Hamiltonian is
\begin{equation}
{\cal H}=-t_x\cos a_xk_x-t_y \cos a_yk_y.\label{HTBdef}
\end{equation}
We set the Fermi level in the middle of the band, $E_{\rm F}=0$, where the open orbits are given by
\begin{equation}
a_xk_x=\pm\arccos\bigl(-\tau\cos a_yk_y\bigr)+2\pi n,\;\;n\in\mathbb{Z},\label{equienergycontour}
\end{equation}
see Fig.\ \ref{fig_dispersion}. 

\begin{figure}[tb]
\centerline{\includegraphics[width=0.8\linewidth]{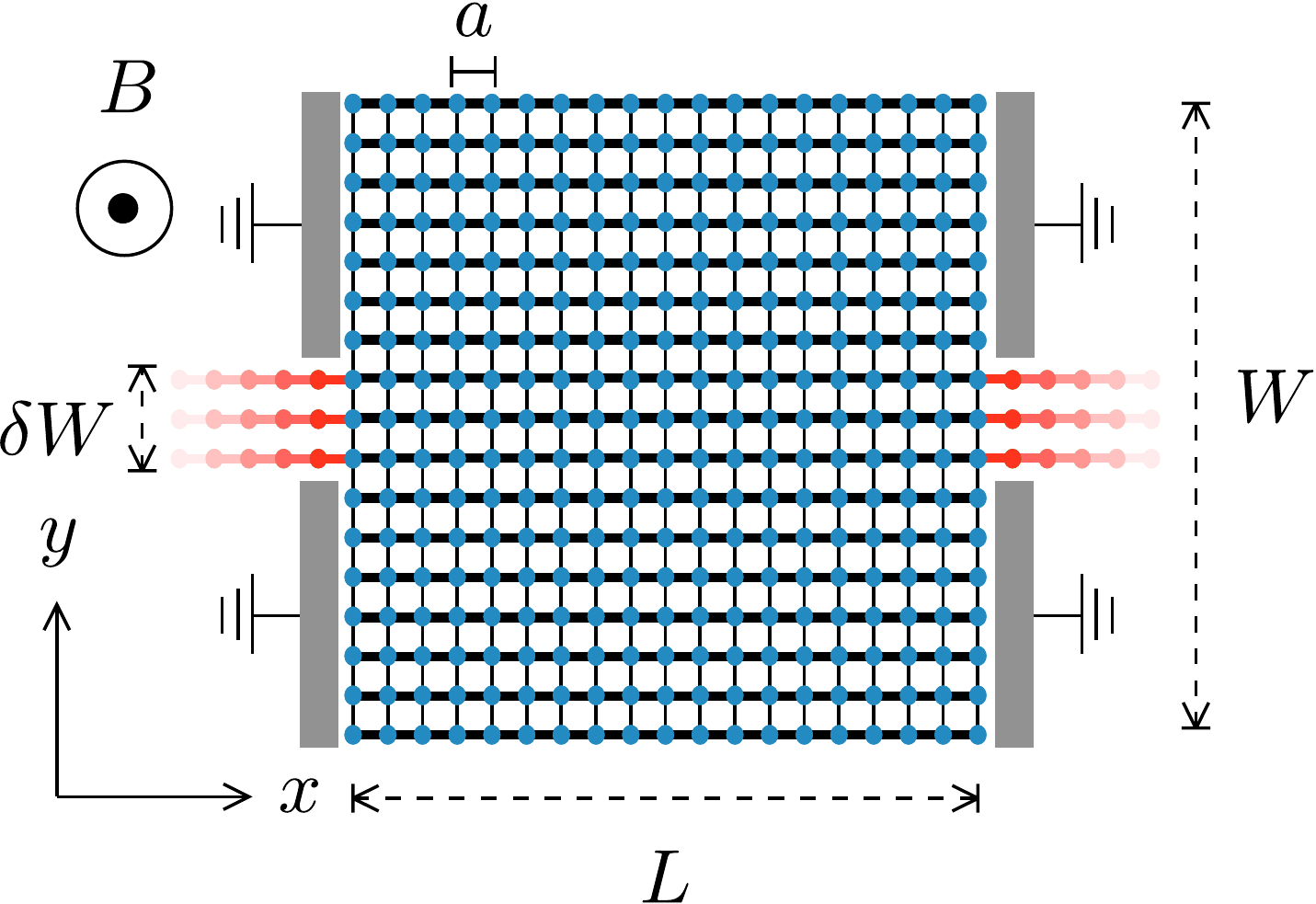}}
\caption{Layout of the tight-binding model, a 2D square lattice with anisotropic hopping energies. Strong and weak bonds are distinguished by thicker and thinner lines. The colors distinguish the conductor (blue), source and drain point contacts (red), and grounded terminals (grey).
}
\label{fig_layout}
\end{figure}

The geometry is shown in Fig.\ \ref{fig_layout}. The conductor has dimensions $L$ in the $x$-direction and $W$ in the $y$-direction. Point contacts (width $\delta W$) at $x=0$ and $x=L$ are a source and drain for electrical current. We implement hard-wall boundary conditions at $|y|=W/2$ (by terminating the lattice) and absorbing boundary conditions at $x=0,L$, $|y|>\delta W/2$ (by attaching ideal leads to ground). The grounded leads are not essential for the magnetoconductance oscillations, they help to improve the resolution by removing a background signal from electrons that are not focused by the lens. 

The point contacts at $x=0,L$, $|y|<\delta W/2$ connect to heavily doped metallic leads, at chemical potential $\mu_{\rm lead}$ much larger than the band width $t_y$ in the conductor. Only a small fraction $t_y/\mu_{\rm lead}$ of the $N\approx \delta W/a$ propagating modes in the leads will couple effectively to the conductor, namely those modes that have transverse momentum small compared to longitudinal momentum. For $t_y/\mu_{\rm lead}\ll 1$ we may thus remove the transverse hoppings in the leads, which are then described by the Hamiltonian \eqref{HTBdef} with $t_y=0$. The perpendicular magnetic field is introduced in the hopping matrix elements via the Peierls substitution.

We use the tight-binding package Kwant \cite{kwant,zenodo} to calculate the scattering matrix of the six-terminal-structure in Fig.\ \ref{fig_layout}. The $N\times N$ transmission matrix $\bm{t}$ from source to drain then gives the conductance $G=(e^2/h)\,{\rm Tr}\,\bm{tt}^\dagger$. 

\begin{figure}[tb]
\centerline{\includegraphics[width=1\linewidth]{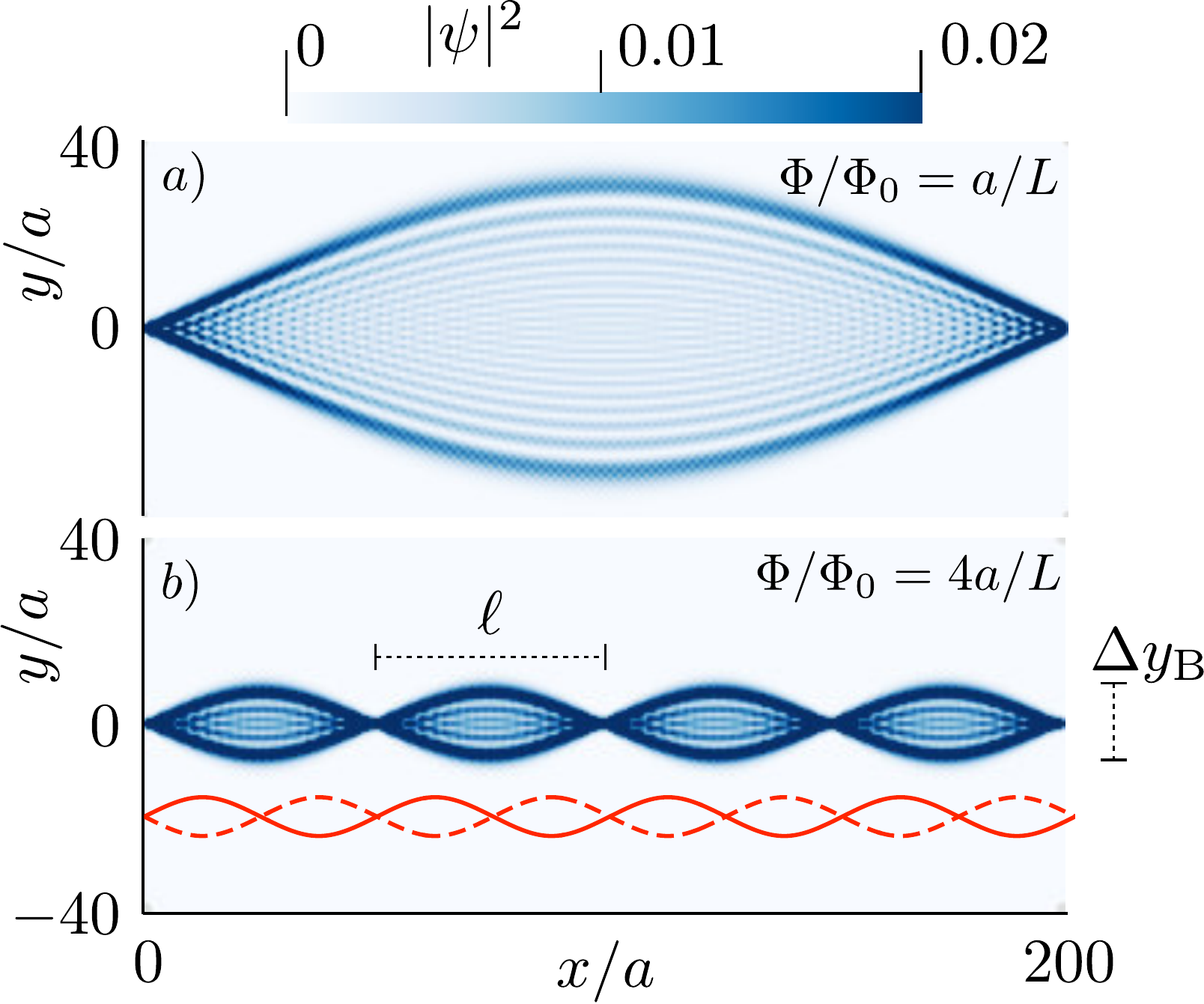}}
\caption{Blue data points: Wave function profile $|\psi(x,y)|^2$ injected into the conductor by a single mode in the lead, for two magnetic fields (corresponding to focal lengths $\ell\equiv a\Phi_0/\Phi=L$ and $\ell=L/4$). The wave function is normalized such that unit current is injected. The red curves in panel b) show two semiclassical orbits, calculated at the same magnetic field value as the breathing mode, to illustrate that the semiclassical orbits oscillate twice as rapidly as the breathing mode envelop.
}
\label{fig_breathingprofile}
\end{figure}

\begin{figure}[tb]
\centerline{\includegraphics[width=0.7\linewidth]{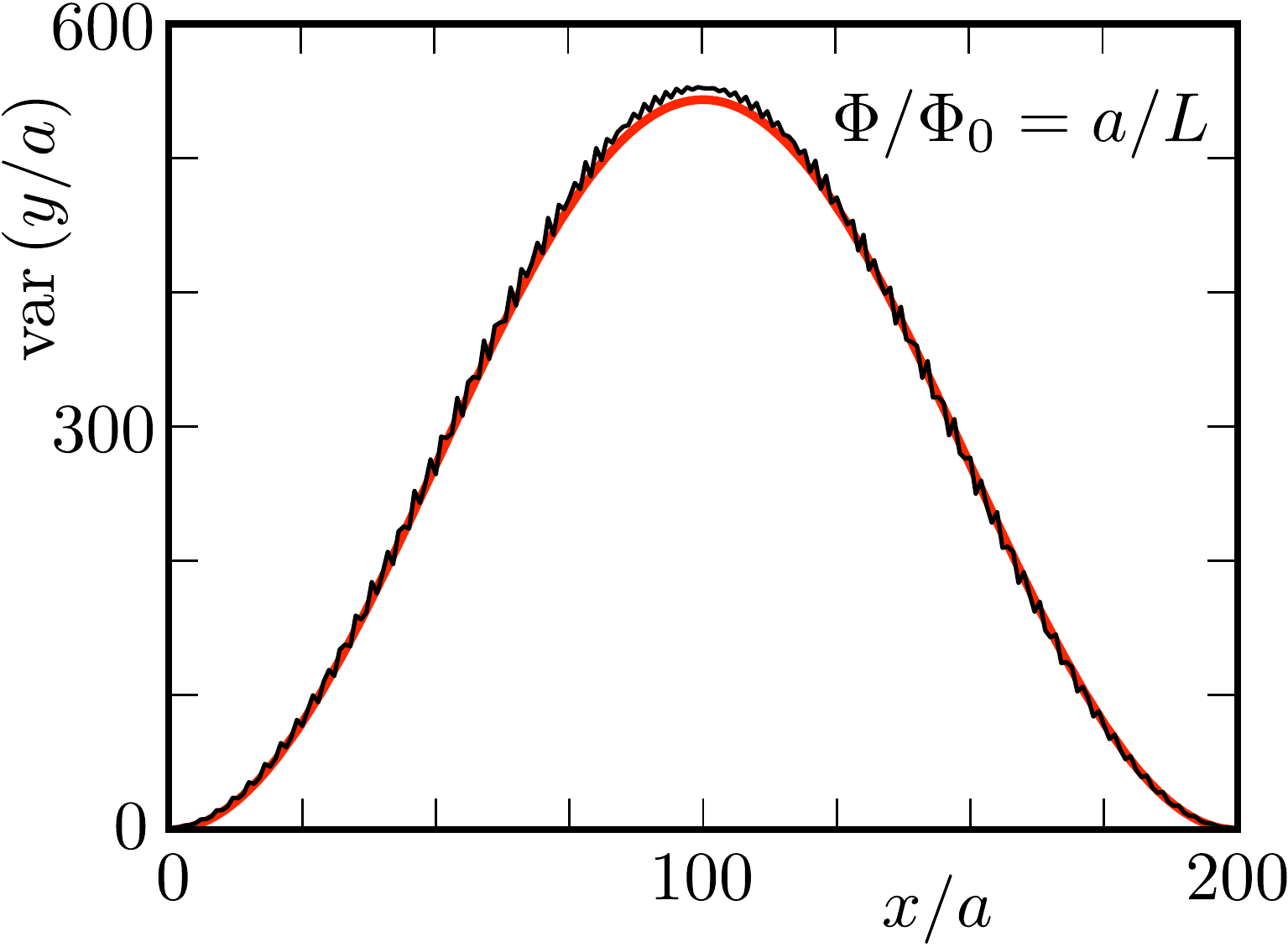}}
\caption{Variance of the spread in the $y$-direction as a function of the distance $x$ from the point source. The smooth red curve is calculated from Eq.\ \eqref{y2analytics}, the black curve with small oscillations is the numerical result from the tight-binding model. The numerical data is obtained by converting the wave function profile in Fig.\ \ref{fig_breathingprofile}a to a normalized intensity profile $\rho_x(y)=|\psi(x,y)|^2/\sum_y|\psi(x,y)|^2$,  and then computing $\sum_y y^2\rho_x(y)$.
}
\label{fig_variance}
\end{figure}

The breathing mode injected into the conductor by a single mode in the lead is shown in Fig.\ \ref{fig_breathingprofile}. It has the expected periodicity of $\Delta x=\ell=a\Phi_0/\Phi$. In Fig.\ \ref{fig_variance} we compare the variance of the spread in the $y$-direction as obtained from the tight-binding model with the result \eqref{y2analytics}. For the open-orbit dispersion we take
\begin{equation}
\varepsilon(k_y)=(\hbar v_x/a)\arccos(-\tau\cos ak_y),\label{eq_dispersion}
\end{equation}
corresponding to one of the two branches in Eq.\ \eqref{equienergycontour}. The agreement is very good, without any adjustable parameter. The small oscillations with periodicity $a$ present in the numerics are due to interference of the two branches of the dispersion relation, which we have neglected in Eq.\ \eqref{eq_dispersion}.  See App.\ \ref{app_oscillations} for a calculation that includes the interference effect.

Because $\hbar\dot{\bm{k}}=e\dot{\bm{r}}\times\bm{B}$, the trajectory $y_c(x)$ of a semiclassical wave packet is obtained from the equi-energy contour $\hbar v_x k_x+\varepsilon(k_y)=0$ upon the transformation $\hbar k_x\mapsto eB_0y$, $\hbar k_y\mapsto -eB_0x$, thus
\begin{equation}
y_c(x)=(eB_0v_x)^{-1}\varepsilon(-eB_0x/\hbar).\label{ycdef}
\end{equation}
A pair of semiclassical orbits is plotted in Fig.\ \ref{fig_breathingprofile}b (red curves), in order to emphasize the fact that the envelope of the breathing mode is not simply the superposition of two semiclassical orbits. Let us study the semiclassical correspondence in more detail.

\section{Semiclassical approximation}
\label{sec_semiclass}

For that purpose we consider (for a state $\psi(x,y)$ normalized to unity) the intensity profile $\rho_x(y)=|\psi(x,y)|^2$ in the weak-field semiclassical regime $\Phi\ll \Phi_0$. We Fourier transform $\rho_x(y)$ with respect to $y$, substitute Eq.\ \eqref{Psixmresult} for $\psi(x,y)$, retain only intensity variations with small wave number $q$, and finally Fourier transform back \cite{note1}:
\begin{align}
\sum_y\,\rho_x(y)e^{iqy}={}&a_y\int_0^{2\pi/a_y} \frac{dk_y}{2\pi}\,\nonumber\\
&\times\exp\bigl(i\omega(x,q+k_y)-i\omega(x,k_y)\bigr)\nonumber\\
=a_y\int_0^{2\pi/a_y}& \frac{dk_y}{2\pi}\,\exp\bigl(iq\partial_{k_y}\omega(x,k_y)+{\cal O}(q^2)\bigr),\\
\Rightarrow \rho_x(y)={}& a_y^2\int_0^{2\pi/a_y} \frac{dk_y}{2\pi}\,\delta(\partial_{k_y}\omega(x,k_y)-y).\label{rhosc}
\end{align}
Now $y$ is treated as a continuous variable (with $\sum_y\mapsto a_y^{-1}\int dy$). 

For a constant magnetic field $B_0$ this can be worked out to
\begin{align}
\rho_x(y)={}&a_y^2 eB_0 v_x\int_0^{2\pi/a_y} \frac{dk_y}{2\pi}\nonumber\\
&\times\delta\bigl[\varepsilon(k_y+eB_0x/\hbar)-\varepsilon(k_y)-eB_0v_xy\bigr].\label{rhoxy1}
\end{align}
In view of Eq.\ \eqref{ycdef} the semiclassical density profile \eqref{rhoxy1} can be rewritten as a superposition of displaced semiclassical orbits,
\begin{align}
\rho_x(y)=(a_y/\ell)\int_0^{\ell} dx_0\,\delta\bigl[y_c(x_0-x)-y_c(x_0)-y\bigr].\label{rhoxy2}
\end{align}

\begin{figure}[tb]
\centerline{\includegraphics[width=0.7\linewidth]{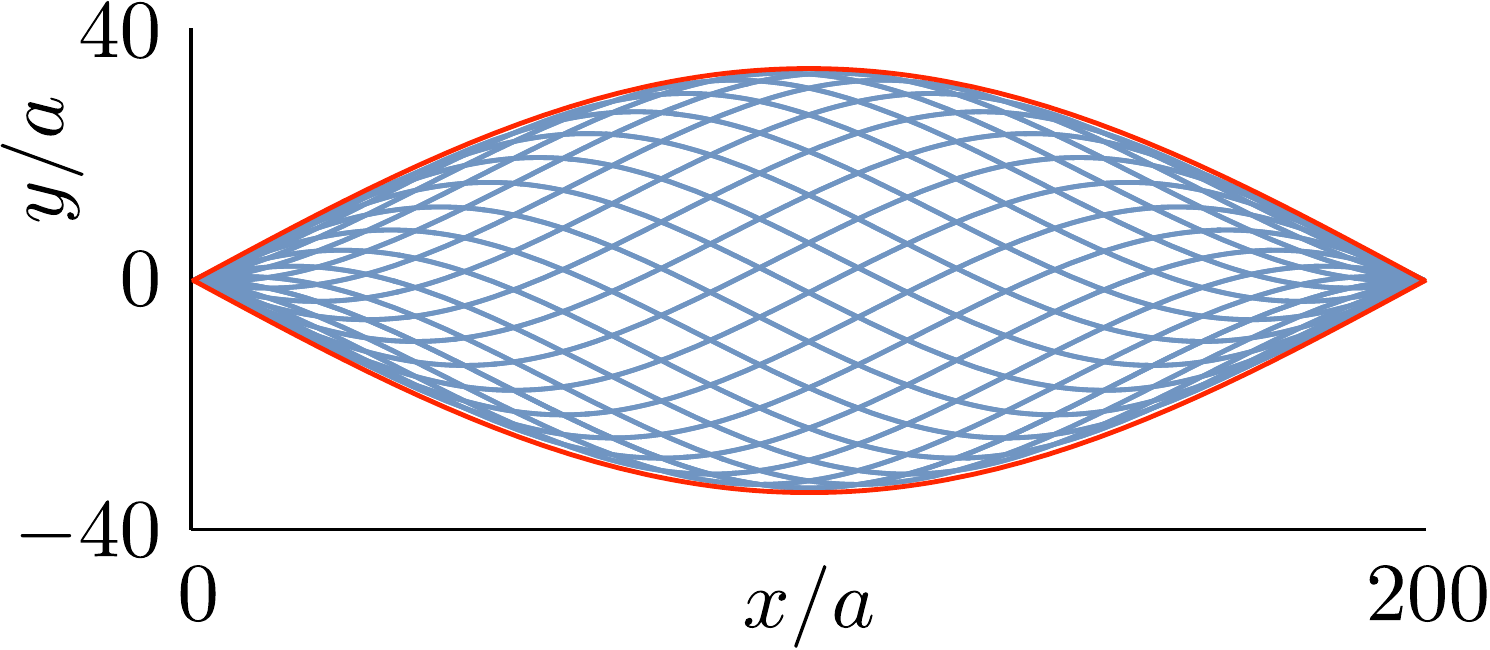}}
\caption{Superposition of semiclassical orbits that satisfy $y_c(x_0-x)-y_c(x_0)-y=0$, with $x_0$ varied between $0$ and $\ell$. The open orbit $y_c(x)$ is given by Eqs.\ \eqref{eq_dispersion} and \eqref{ycdef}, the parameters are those of Fig.\ \ref{fig_breathingprofile}a. The caustic is indicated in red.
}
\label{fig_caustic}
\end{figure}

In Fig.\ \ref{fig_caustic} we have plotted this superposition for the same parameters as in the tight-binding simulation of Fig.\ \ref{fig_breathingprofile}a. The profiles match very well. The semiclassical calculation identifies the envelope as a caustic: an accumulation of classical trajectories with an infinite density, regularized by the finite wave length in the quantum calculation.

Eq.\ \eqref{rhoxy1} allows for a semiclassical estimate for the amplitude of the breathing mode: Since $\rho_x(y)\equiv 0$ for all $x$ when $|y|>(eB_0v_x)^{-1}\max_{k_1,k_2}|\varepsilon(k_1)-\varepsilon(k_2)|\equiv (eB_0/\hbar)^{-1}\Delta k_x$ or, equivalently, when $|y|>\max_{x_1,x_2}|y_c(x_1)-y_c(x_2)|\equiv \Delta y_c$, we arrive at the relation
\begin{equation}
\Delta y_{\text B} =2 \Delta y_c=2(\hbar/eB_0) \Delta k_x 
\end{equation}
between the amplitude $\Delta k_x$ of the open orbit in momentum space, on the one hand, and the amplitudes $\Delta y_{\text B}$ and $\Delta y_c$ of breathing mode and semiclassical orbit in real space, on the other hand. 

The ratio $R=\Delta y_{\text B}/\ell=(a_y/\pi)\Delta k_x$ is a magnetic field independent characteristic of the open orbit. For the anisotropic dispersion \eqref{HTBdef} one has
\begin{equation}
R\equiv\Delta y_{\text B}/\ell=(2a_y/\pi a_x)\arcsin (t_y/t_x).
\end{equation}
The ratio equals 1/3 for the parameters in Fig.\ \ref{fig_breathingprofile} ($a_y=a_y$, $t_x=2t_y$).

\section{Magnetoconductance oscillations}
\label{sec_magnetocond}

In the double point-contact geometry of Fig.\ \ref{fig_layout} the breathing mode manifests itself as a conductance peak when the point contact separation $L$ is an integer multiple of the period $\ell$. This is the magnetoconductance oscillation studied in the context of twisted bilayer graphene in Ref.\ \onlinecite{Vak22}. The magnetic field periodicity is
\begin{equation}
\Delta B=\frac{h}{eaL}.\label{DeltaBperiod}
\end{equation}

\begin{figure}[tb]
\centerline{\includegraphics[width=0.8\linewidth]{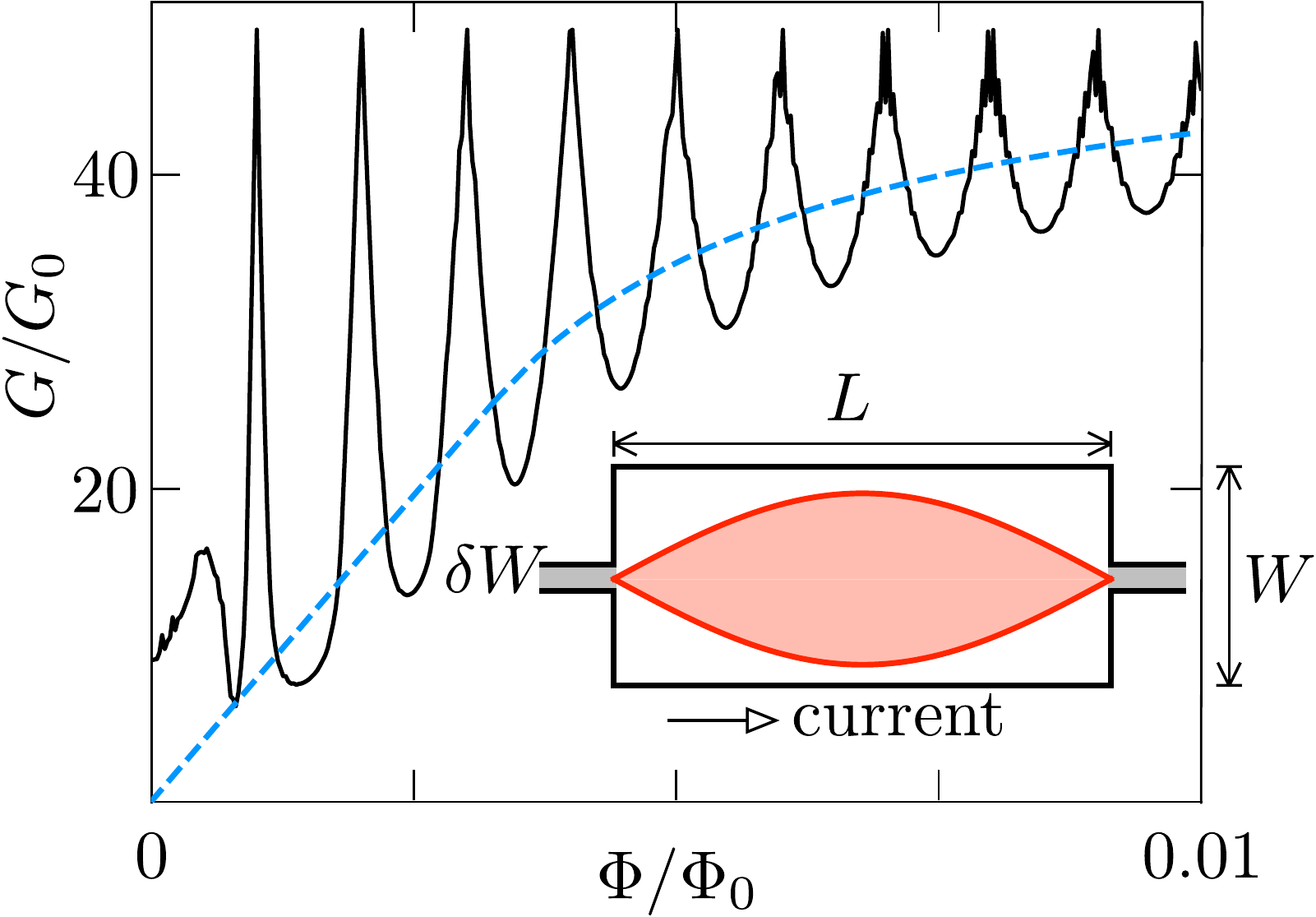}}
\caption{Conductance $G$ (in units of the conductance quantum $G_0=e^2/h$, per spin degree of freedom), as a function of magnetic field $B=\Phi/a^2$, computed from the tight-binding model in the point contact geometry shown in the inset (parameters $L/a=1000$, $W/a=440$, $\delta W/a=51$). The breathing mode at the first conductance peak is shown in red. The periodicity of the oscillations is $\Delta\Phi/\Phi_0=a/L=10^{-3}$. Full refocusing of the breathing mode without any backscattering would give a conductance peak of $NG_0$ with $N=51$ injected modes. The blue dashed curve is the calculated decay  \eqref{GminGmax} of the amplitude of the conductance oscillations.
}
\label{fig_conductance}
\end{figure}

A simulation of the tight-binding model in Fig.\ \ref{fig_conductance} shows the effect. The amplitude of the oscillations decays with increasing field because the point contact width $\delta W$ is no longer able to resolve the decreasing amplitude $\Delta y_{\rm B}$ of the breathing mode. In terms of the dimensionless parameter $\xi(B)=(1/R)(\delta W/a_y)(\Phi/\Phi_0)$ we calculate that the ratio $G_{\rm min}/G_{\rm max}$ of the minima and maxima of the conductance oscillations follows the curve\cite{note3}
\begin{equation}
\frac{G_{\rm min}}{G_{\rm max}}=\begin{cases}
\xi(B)&\text{if}\;\;\xi(B)<1/2,\\
1-\tfrac{1}{4}\xi(B)^{-1}&\text{if}\;\;\xi(B)>1/2.
\end{cases}\label{GminGmax}
\end{equation}
This agrees quite nicely with the numerics (blue curve in Fig.\ \ref{fig_conductance}, without any fit parameter.

\begin{figure}[tb]
\centerline{\includegraphics[width=0.8\linewidth]{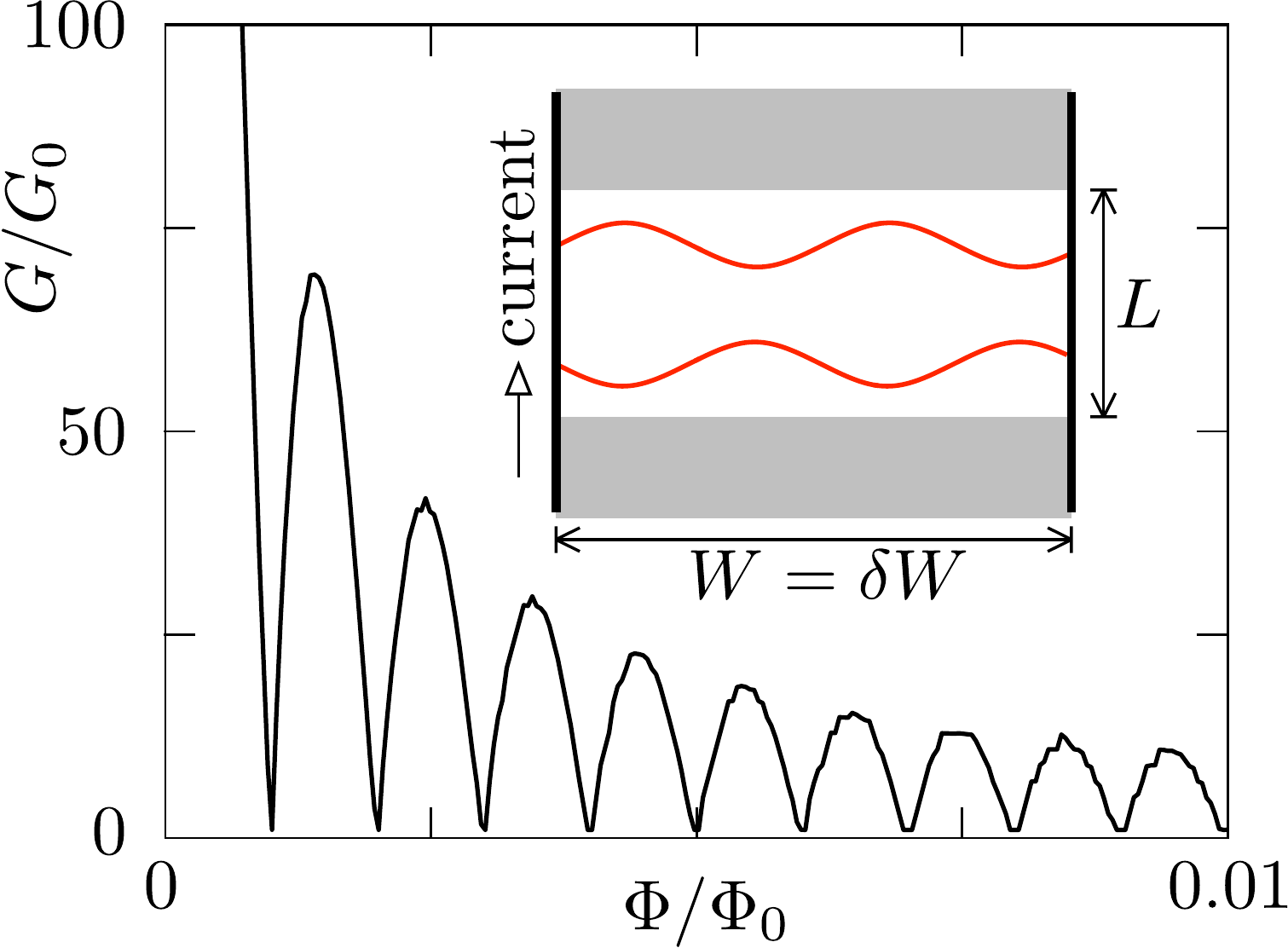}}
\caption{Same as Fig.\ \ref{fig_conductance}, but now with the current flowing perpendicularly to the open orbits (a few are shown as red trajectories; parameters $W/a=1000$, $L/a=440$, no point contacts, $\delta W=W$). The conductance has a minimum when an open orbit fits in the width of the conductor, so when $B=n\Phi_0/aW$, $n=1,2,\ldots$.
}
\label{fig_Pippard}
\end{figure}

To make contact with Refs.\ \onlinecite{Put20,Vil20}, we note that magnetoconductance oscillations with the same period \eqref{DeltaBperiod} --- upon exchange of $L$ by $W$ --- can be observed without any point contacts, so without focusing of wave profiles. Instead of a current flowing along the open orbit the current should then flow perpendicularly to the open orbit, see Fig.\ \ref{fig_Pippard}. This is the geometry first studied by Pippard \cite{Pip65,note2}, to explain conductance oscillations with period $\Phi_0/aW$ in cadmium \cite{Mun65,Mun68}. We refer to Refs.\ \onlinecite{Put20,Vil20} for a comprehensive theory and experiment on these magnetoconductance oscillations. Note that magnetic lensing plays no role in the Pippard geometry, one needs the spatial resolution of a point contact to excite a breathing mode.

\section{Conclusion}
\label{conclude}

In summary, we have presented a magnetic lensing effect with an unusually long focal length, set by the Bragg momentum $\hbar/a$ rather than the Fermi momentum $p_{\rm F}$. At a field of 1~T and for a lattice constant $a=0.5~\text{nm}$ the focal length $\ell=h/eBa\approx 8~\mu{\rm m}$ --- an order of magnitude larger than in semiconductor electron focusing experiments \cite{Bee91}. Magnetic focusing is an effective way to study scattering processes \cite{Gup21} and in clean systems a large focal length would be an advantage.

The quantum mechanical origin of the focusing effect, Bragg reflection at Brillouin zone boundaries, does not imply that the magnetic lens needs long-range phase coherence --- the breathing mode only requires phase coherence on the scale of the lattice constant. We note the contrast with the Aharanov-Bohm effect, where a magnetoconductance oscillation with period $h/eS$ would require phase coherence over distances of order $\sqrt S$.  The oscillation period \eqref{DeltaBperiod} has $S=aL$ but only requires phase coherence over a length $a$, irrespectively of how large $L$ might be.

We have applied the general theory to a simple model of an anisotropic dispersion, appropriate for the layered material (delafossites) studied in Refs.\ \onlinecite{Put20,Vil20} (with a ratio $\tau\simeq 10^{-2}$ between in-plane and out-of-plane hopping energies, and mean free paths of $20\,\mu{\rm m}$ \cite{Tak13}). For such a strong anisotropy the open orbits in the Brillouin zone are essentially decoupled from each other, allowing for closed-form expressions for the breathing mode in the fully quantum regime, Eq.\ \eqref{Psixmresult}, and in the semiclassical approximation, Eq.\ \eqref{rhoxy2}. 

More complicated band structures would allow for multiple open orbits coupled by magnetic breakdown. The magnetic lens may then exhibit a complex pattern of caustics, one example (relevant for twisted bilayer graphene \cite{Vak22}) is analysed in App.\ \ref{app_multiple}.

\acknowledgments

This project has received funding from the Netherlands Organization for Scientific Research (NWO/OCW) and from the European Research Council (ERC) under the European Union's Horizon 2020 research and innovation programme.

\appendix

\section{Calculation of the interference oscillations in the root-mean-square displacement}
\label{app_oscillations}

The tight-binding model calculation in Fig.\ \ref{fig_variance} shows small oscillations on the scale of the lattice constant, which are absent in the analytical curve. To include these, we consider both branches of the equi-energy contour \eqref{equienergycontour}. These produce two open-orbit dispersions $\pm\varepsilon(k_y)$, with two corresponding wave function profiles $\psi_\pm$. With reference to Eq.\ \eqref{Psixmresult} we have $\psi_+(x,y)=\psi(x,y)$ and $\psi_-(x,y)=\psi^\ast(x,-y)$.

\begin{figure}[tb]
\centerline{\includegraphics[width=0.7\linewidth]{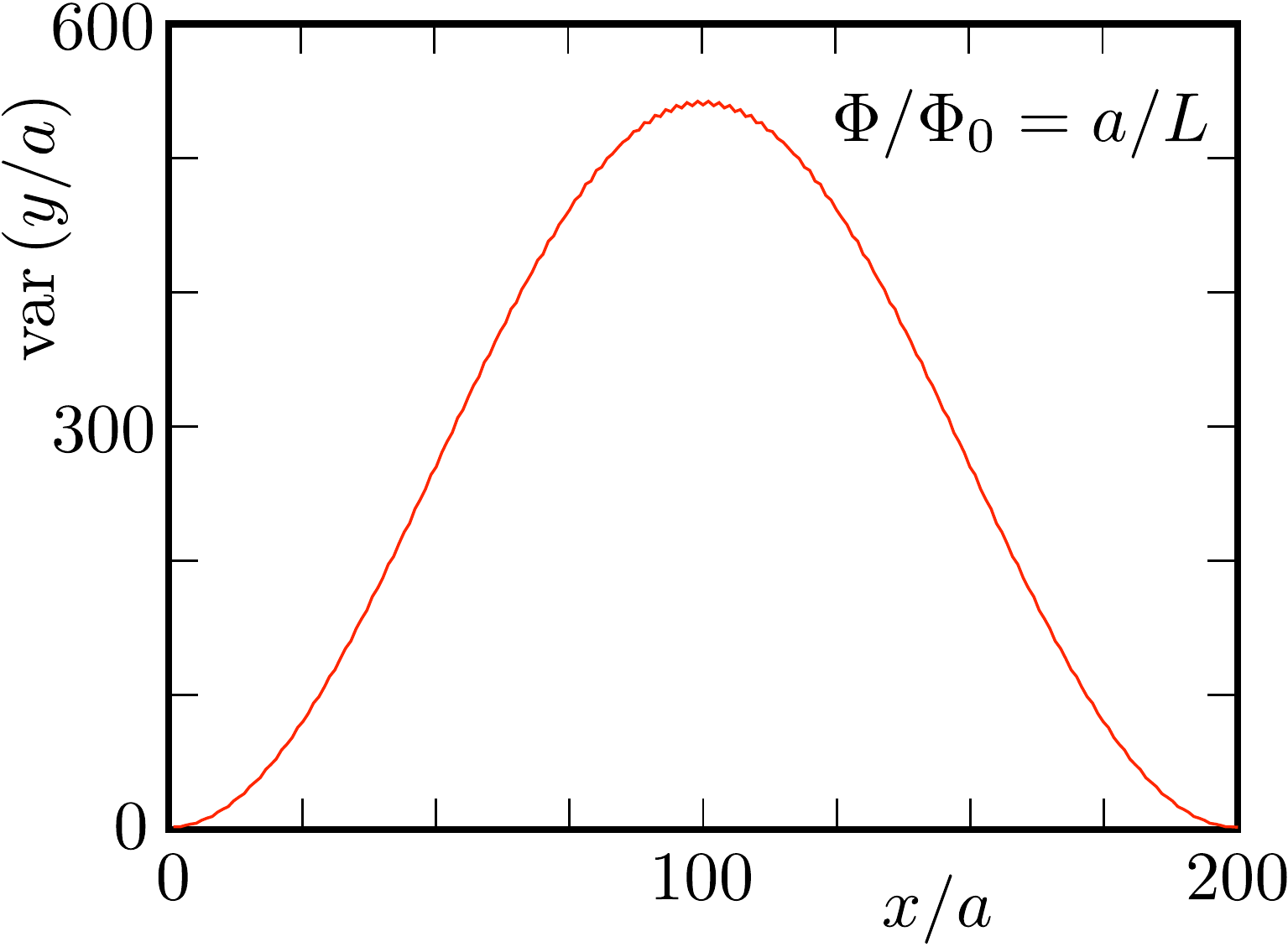}}
\caption{Variance of the spread in the $y$-direction as a function of the distance $x$ from the point source, calculated from Eq.\ \eqref{oscillation}. This figure can be compared with Fig.\ \ref{fig_variance}, where the interference oscillations are neglected.
}
\label{fig_variance2}
\end{figure}

We take an equal weight superposition $2^{-1/2}(\psi_++\psi_-)$. The average displacement remains equal to zero, the mean square displacement becomes
\begin{equation}
	\left\langle y^2\right\rangle_x=a_y \int_0^{2 \pi / a_y} \frac{d k_y}{2 \pi}  \left(\partial_{k_y}\omega(x,k_y)\right)^2 [1 - \cos 2\omega(x,k_y)] .\label{oscillation}
\end{equation}
The result, see Fig.\ \ref{fig_variance2}, has oscillations with a somewhat smaller amplitude than in the numerics of Fig.\ \ref{fig_variance}, but the periodicity agrees nicely.

\section{Magnetic lens for multiple coupled open orbits}
\label{app_multiple}

\begin{figure}[tb]
\centerline{\includegraphics[width=0.9\linewidth]{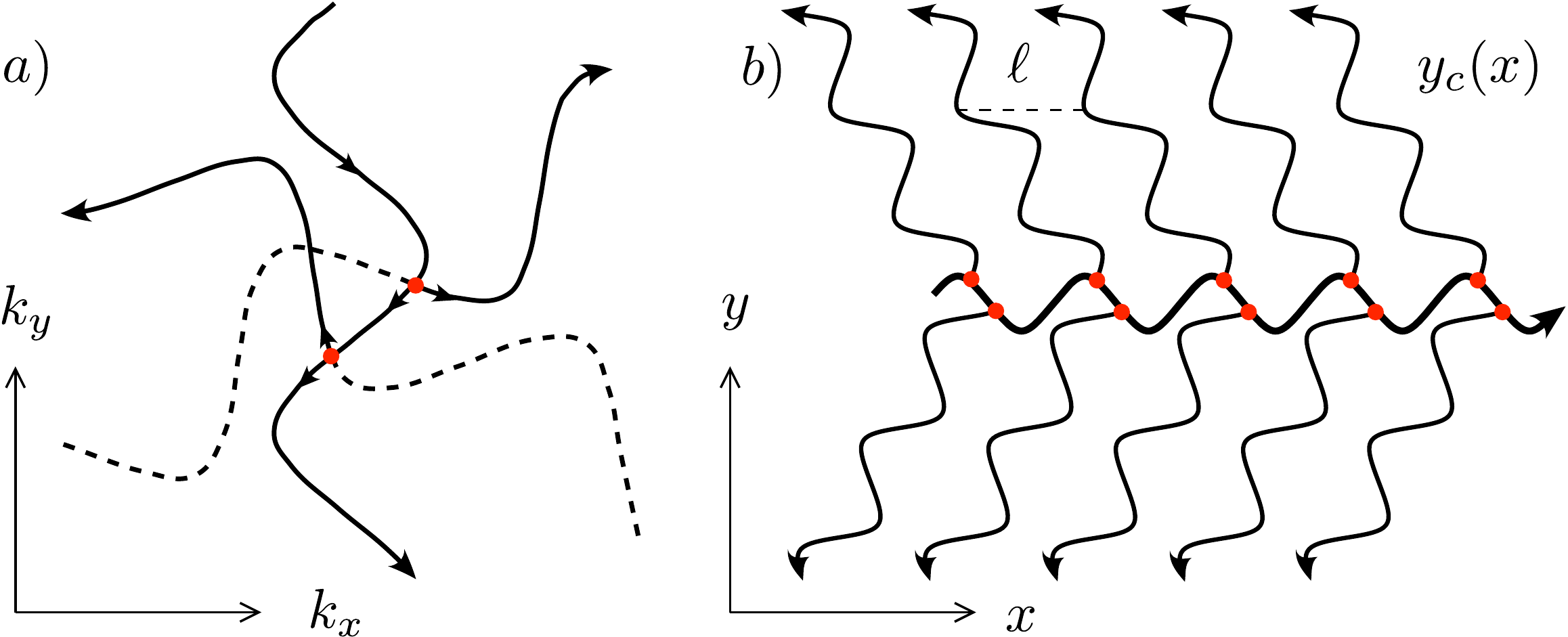}}
\caption{Panel \textit{a}: Equi-energy contours in momentum space consisting of three sets of open orbits, at relative orientation of $120^{\circ}$. The arrows indicate the direction of motion in a magnetic field. The solid contours produce, upon rotation by $90^\circ$, the multi-branched real-space trajectory $y_c(x)$ shown in panel \textit{b}. A trajectory initially moving in the $+x$ direction branches out into the $-x$ direction at the intersection points indicated by red dots. Higher order branch-outs are not considered, these would contribute with reduced amplitude.
}
\label{fig_orbits}
\end{figure}

\begin{figure}[tb]
\centerline{\includegraphics[width=0.8\linewidth]{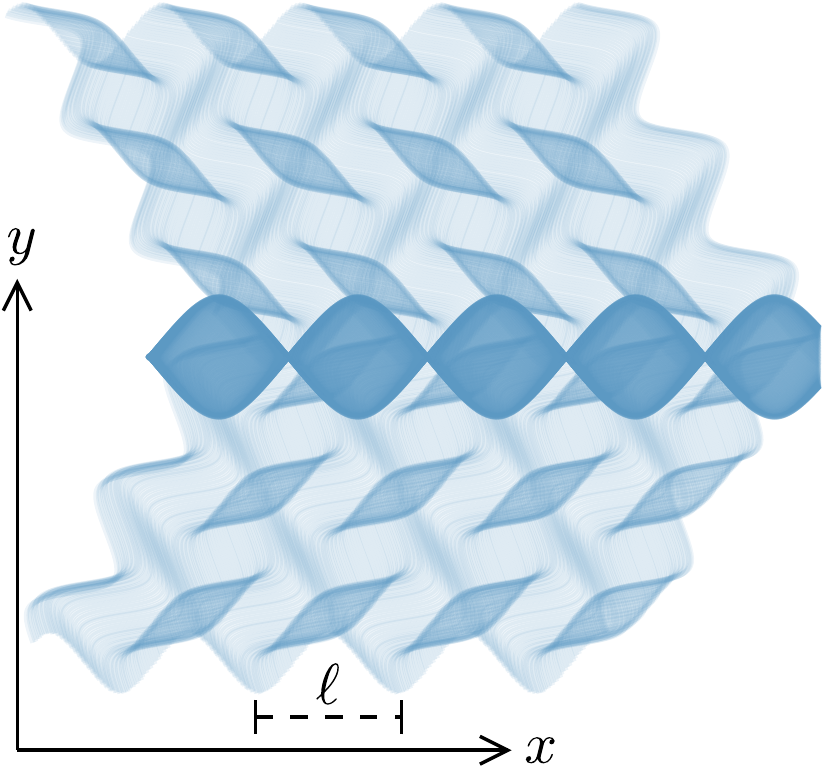}}
\caption{Superposition of the semiclassical orbits $y_c(x)$ from Fig.\ \ref{fig_orbits}b that satisfy $y_c(x_0-x)-y_c(x_0)-y=0$, with $x_0$ varied between $0$ and $\ell$.
}
\label{fig_caustic2}
\end{figure}

In the main text we considered the magnetic lens that results from a single open orbit in the Brillouin zone. As a more complicated example, we show in Fig.\ \ref{fig_orbits}a the equi-energy contours of minimally twisted bilayer graphene \cite{Beu20}, with three open orbits at a relative orientation of $120^\circ$. At an intersection an electron can switch from one orbit to the other, a process known as magnetic breakdown. The corresponding multi-branched classical trajectory $y_c(x)$ is shown in Fig.\ \ref{fig_orbits}b. If we now apply the semiclassical formula \eqref{rhoxy2} we obtain the complex pattern of caustics shown in Fig.\ \ref{fig_caustic2}.

In Ref.\ \onlinecite{Vak22} a fully quantum mechanical calculation was presented for the wave function profile. The semiclassical calculation well reproduces the qualitative features. Notice in particular that the side branches at an orientation of $120^\circ$ are not simply copies of the main breathing mode. There is an extinction of the amplitude between two oscillations, which one might have suspected to be an interference effect. Instead it can be fully reproduced from a trajectory description.


\begin{thebibliography}{99}
\bibitem{note1}  We note that the expression for the variance $\langle y^2\rangle_x=a_y^{-1}\int dy\, y^2\rho_x(y)$ which follows from the semiclassical density profile \eqref{rhosc} is \textit{identical} to the result \eqref{y2exact} which we obtained without making the semiclassical approximation. We have no more fundamental explanation for this correspondence. It does not carry over to higher moments.
\bibitem{note3} Eq.\ \eqref{GminGmax} follows from a calculation of the fraction of the injected breathing mode that can enter the opposite point contact when $L=(n+1/2)\ell$, so at the conductance minimum. This fraction equals $G_{\rm min}/G_{\rm max}$ and is given by the integral $(\delta W\Delta y_{\rm B})^{-1}\int_{-\delta W/2}^{\delta W/2}dx\,\bigl(\min\bigl[\delta W/2,x+\Delta y_{\rm B}/2\bigr]-\max\bigl[-\delta W/2,x-\Delta y_{\rm B}/2\bigr]\bigr)$.
\end{thebibliography}
\end{document}